\documentclass[twocolumn,tighten,trackchanges]{aastex631}
\usepackage{amsmath}
\hypersetup{linkcolor=blue,citecolor=blue,filecolor=cyan,urlcolor=blue}
\shorttitle{Cross-Equatorial Meridional Flows}
\shortauthors{Sen et al. }
\graphicspath{{./}{figures/}}
\begin{document}

\title{Hemispheric Magnetic Asymmetry and Cross-Equatorial Circulation Cells within the Sun's Near-Surface Shear Layer}

\author[0000-0003-2694-3288]{Anisha Sen} 
\email{anisha.sen@iiap.res.in}
\affiliation{Indian Institute of Astrophysics, II Block, Koramangala, Bengaluru 560 034, India}
\affiliation{Pondicherry University, R.V.Nagar, Kalapet, Puducherry 605014, India}

\author[0000-0003-0003-4561]{S.P. Rajaguru}
\email{rajaguru@iiap.res.in}
\affiliation{Indian Institute of Astrophysics, II Block, Koramangala, Bengaluru 560 034, India}


\author[0000-0002-2632-130X]{Ruizhu Chen}
\affiliation{W. W. Hansen Experimental Physics Laboratory, Stanford University, Stanford, CA 94305-4085, USA}

\author[0000-0002-6308-872X]{Junwei Zhao}
\affiliation{W. W. Hansen Experimental Physics Laboratory, Stanford University, Stanford, CA 94305-4085, USA}

\author[0000-0003-1860-3697]{Shukur Kholikov}
\affiliation{National Solar Observatory, Boulder CO, USA}









\begin{abstract}

Using time-distance helioseismic measurements of meridional flow in the near-surface shear layer over a period 
of 14 years starting from May 2010, we probe the depth structure and evolution of its cross-equatorial part. 
We confirm that the hemispheric magnetic asymmetry determines the amplitude and direction of such flows. Additionally,
we find that these flows turn over and change direction at depths below 0.97$R_{\sun}$, forming circulation cells with 
lifetimes dictated again by the hemispheric magnetic imbalance, which is dominated by the occurrences of large sunspots. 
We also examine connections between cross-equatorial magnetic flux plumes and the flows, and discuss their implications 
for the equatorial flux cancellation/submergence and the poleward transport of flux.  
 
\end{abstract}



\section{Introduction} 
\label{sec:intro}

Global scale meridional flow \citep{1979SoPh...63....3D,2010Sci...327.1350H,2013ApJ...774L..29Z,2014ApJ...784..145K,
rajaguru2015meridional,2017ApJ...849..144C} is well recognised as a key player in the magnetic flux transport processes (see e.g.\ 
\citet{wang1989magnetic,2006ApJ...649..498D,2023SSRv..219...31Y}) on the Sun. In the near-surface layers, it carries flux from 
decaying active regions, typically the trailing-polarity fields of tilted bipolar regions, toward the poles, where they cancel 
oppositely oriented fields of the previous solar cycle and drives the reversal of the Sun’s polar magnetic fields 
\citep{wang1989magnetic,mackay2012sun}. While we still lack a complete understanding of the origin and maintenance of 
meridional circulation on the Sun, on average, it is modelled as a hemisphere-antisymmetric flow system with a vanishing 
meridional component at the equator. However, in reality, on time scales of the life-times of active region complexes, which 
have hemispheric asymmetry, prominent cross-equatorial flows are observed \citep{komm2022subsurface}. Such flows are thought 
to play a significant role in transporting opposite-polarity magnetic flux across the equator, facilitating magnetic flux 
cancellation \citep{norton2014hemispheric}. Some studies indicate that the total flux cancelled at the equator directly 
correlates with the net flux transported to the poles \citep{2012A&A...548A..57C,bisoi2020new}. Earlier studies had linked 
substantial deviations in the average meridional flow at the equator to minor systematic errors in telescope alignment at 
single-site, ground-based observations \citep{komm1993meridional}. However, modern helioseismic instruments have undergone 
rigorous alignment verification, including validation through planetary transits for Global Oscillation Network Group 
(GONG)\citep{toner2001absolute, toner2004first} and the Helioseismic and Magnetic Imager (HMI) onboard the Solar Dynamics 
Observatory (SDO) \citep{couvidat2016observables,hoeksema2018orbit}. Such precise calibrations ensure reliable measurement of 
small meridional flow variations near the equator. Through ring-diagram analysis of Dopplergrams from the Michelson Doppler 
Imager (MDI) Dynamics Program, the GONG and HMI, \citet{komm2022subsurface} reported significant cross-equatorial meridional 
flows toward the hemisphere with larger magnetic flux, to depths down to 10 Mm. Using solar-cycle-long time-distance 
helioseismic measurements of meridional flows in the Sun’s near-surface shear layer (NSSL), \citet{2025ApJ...984L...1S} 
identified that near-surface inflows toward active latitudes are part of a localized circulation, which has an outflow away 
from active latitudes at depths of approximately 0.97$R_{\sun}$ (20 Mm). These authors also showed that such active region 
flows, under the action of Coriolis force, explain the depth profile of deviations in the radial gradient of rotation measured 
in global helioseismic studies \citep{antia2022changes}.


Investigating the dynamics of active-region driven circulation cells near or across the equator 
\citep{2004SoPh..220..371H,hindman2009subsurface,braun2019flows} and their variations in response to 
hemispheric asymmetry of magnetic flux is essential to understanding the flux cancellation process and the
resulting global evolution of solar-cycle magnetic field \citep{2012A&A...548A..57C}. 
In this paper, we examine the depth profiles of cross-equatorial flows within the NSSL
using time-distance helioseismic measurements and investigate the extent to which these flows are 
influenced by large-scale inflows toward active regions, offering insights into their role in solar dynamics and 
internal flow variations. The rest of the paper is structured as follows: Section \ref{sec: data} outlines the 
data utilized and a description of the analysis technique. Section \ref{sec: result} presents our findings, and 
Section \ref{sec:conclusion} discusses the implications of our results.

\section{Data and Analysis Procedure} \label{sec: data}

The data and analysis procedure are the same as those used and explained by \citet{2025ApJ...984L...1S}. Briefly, using 
identically processed helioseismic data from the space-borne Helioseismic and Magnetic Imager (HMI) 
\citep{scherrer2012helioseismic} aboard NASA’s Solar Dynamics Observatory (SDO) and from the ground-based Global Oscillation 
Network Group (GONG), we perform time-distance helioseismology \citep{duvall1993time} to measure meridional flows 
\citep{rajaguru2015meridional} within the NSSL. The data covers a 14-year period from May 2010 to April 2024.
Our measurements are at a binned-down spatial resolution of 0.36 degrees per pixel for both HMI and GONG.
To take care of the surface magnetic effect in flow measurements \citep{liang2015effects,2017ApJ...849..144C}, 
we mask out active regions in input Doppler data that exceed a threshold of 40 G in the 0.36 deg/pix resolution HMI LOS 
magnetograms. We further note that we recover meridional flows from the inversions for the stream function, which satisfies 
the continuity equation and thus the mass conservation constraint is built into the inversion scheme
\citep{rajaguru2015meridional}.

We also utilize local time-distance helioseismic inversions for horizontal velocity fields \citep{2012SoPh..275..375Z}, 
available through the JSOC time-distance helioseismology pipeline{\footnote{\textcolor{blue} 
{http://jsoc.stanford.edu/data/timed/}}}. Additionally, we analyse HMI line-of-sight magnetograms to investigate the temporal 
and latitudinal variations of magnetic flux. To compare time-latitude profiles of cross-equatorial flows with sunspot 
distributions, we use data from the National Oceanic and Atmospheric Administration (NOAA) Solar Region Summary.

\graphicspath{ {newfigures/} }

\section{Results} \label{sec: result}

\subsection{Cross-Equatorial Flows: Structure, Evolution, and Active Region Connections}
\label{subsec: res1}

\begin{figure*}[!htbp]
\gridline{\fig{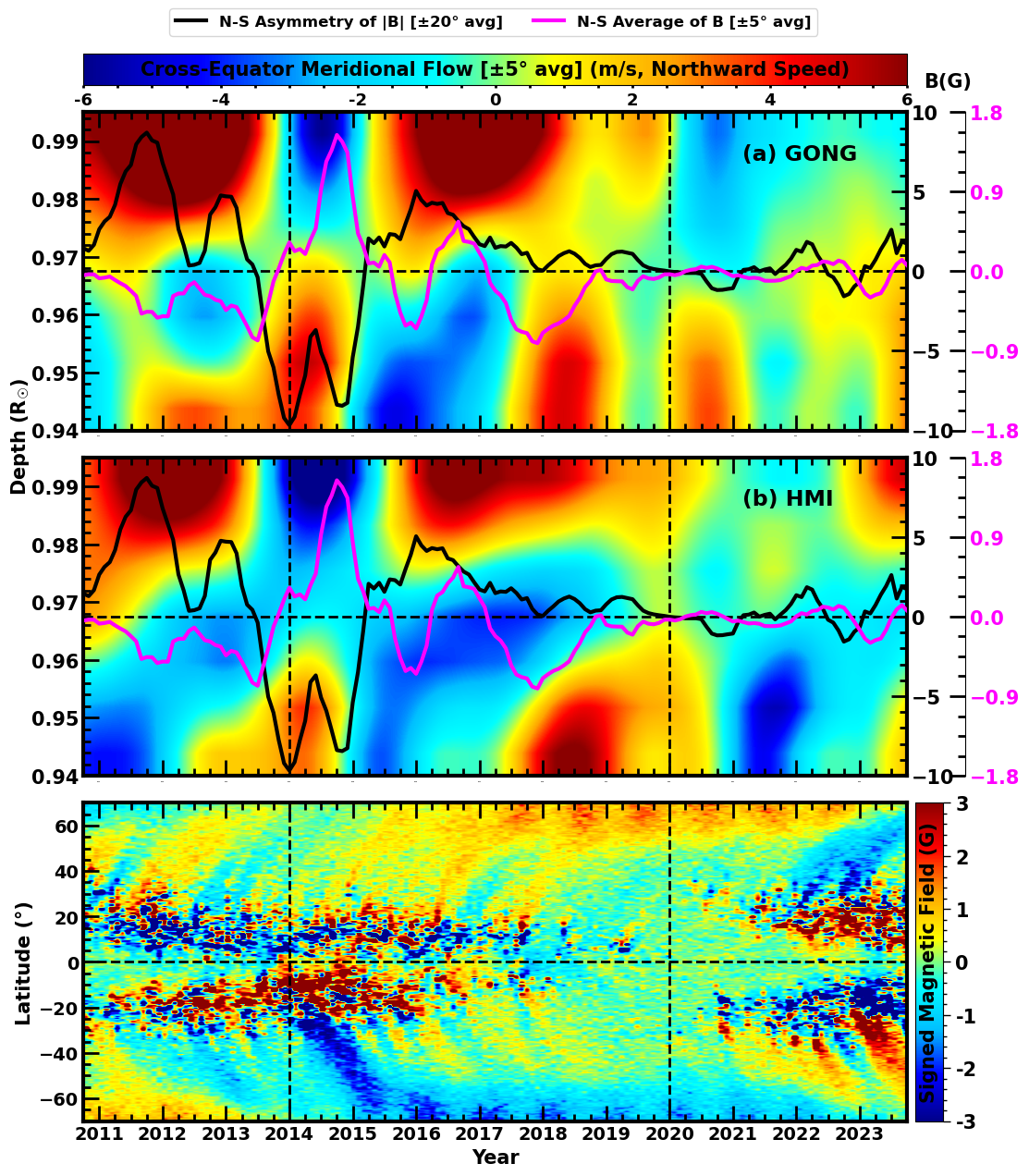}{0.70\textwidth}{}          }

\caption{Time-depth profiles of meridional flow, $U_{\theta}$, averaged over $\pm 5^{o}$ across the equator, are shown in the 
top two panels that compare measurements from GONG and HMI. Positive (negative) values represent 
northward (southward) flows, and the depth range covers the whole of the near-surface shear layer. The bottom panel shows the 
magnetic butterfly diagram derived from the HMI LOS magnetic field, denoted simply as B, for the same period. The overplotted 
black and pink curves, with right y-axes, are the time-variations N-S asymmetry in the absolute hemispheric magnetic field 
averaged over active latitudes ($\pm 20^{o}$), $< \vert$B$\vert_{N}>$ - $<\vert$B$\vert_{S}>$, and N-S average of signed 
magnetic field over the equator ($\pm 5^{o}$), respectively. Both curves have been smoothed using a 6-month running average.}

\label{fig:1}
\end{figure*}

A primary objective in this work is to probe the depth structure of cross-equatorial meridional flows. 
Time-depth profiles of equator-crossing part of meridional flow, $U_{\theta}$, averaged over $\pm 5^{o}$ across the 
equator and covering the whole depth range of the NSSL, are shown in the top and middle panels of Figure \ref{fig:1} for the
GONG and HMI measurements, respectively. The sign convention of positive values for northward flow in both hemispheres is used,
and a 12-month running average has been applied for the flow measurements. A very good agreement of flow measurements
from the two independent data sources is clear.

To study the connections to magnetic flux, we generate the magnetic butterfly diagram using the HMI LOS magnetic fields, 
denoted simply as B, and it is shown in the bottom panel of Figure \ref{fig:1}. The hemispheric asymmetry in the magnetic field 
is estimated by taking the difference between the absolute magnetic field averaged over active latitudes ($0 - 20^{\circ}$) 
in the north and south, B$_{asym}$=$< \vert$B$\vert_{N}>$ - $<\vert$B$\vert_{S}>$. We also calculate the N-S average of the 
signed magnetic field over the equator ($\pm 5^{o}$), B$_{eq}$, which represents the cross-equatorial flux plumes 
\citep{2013A&A...557A.141C,bisoi2020new}, to examine their connections to the cross-equatorial flows. Both these quantities 
are overplotted as black and pink curves in Figure \ref{fig:1}, with values in the right Y-axes. We discuss the connections 
between cross-equatorial flows and flux plumes (B$_{eq}$) in the next sub-section.

\begin{figure*}[!htbp]
\gridline{\fig{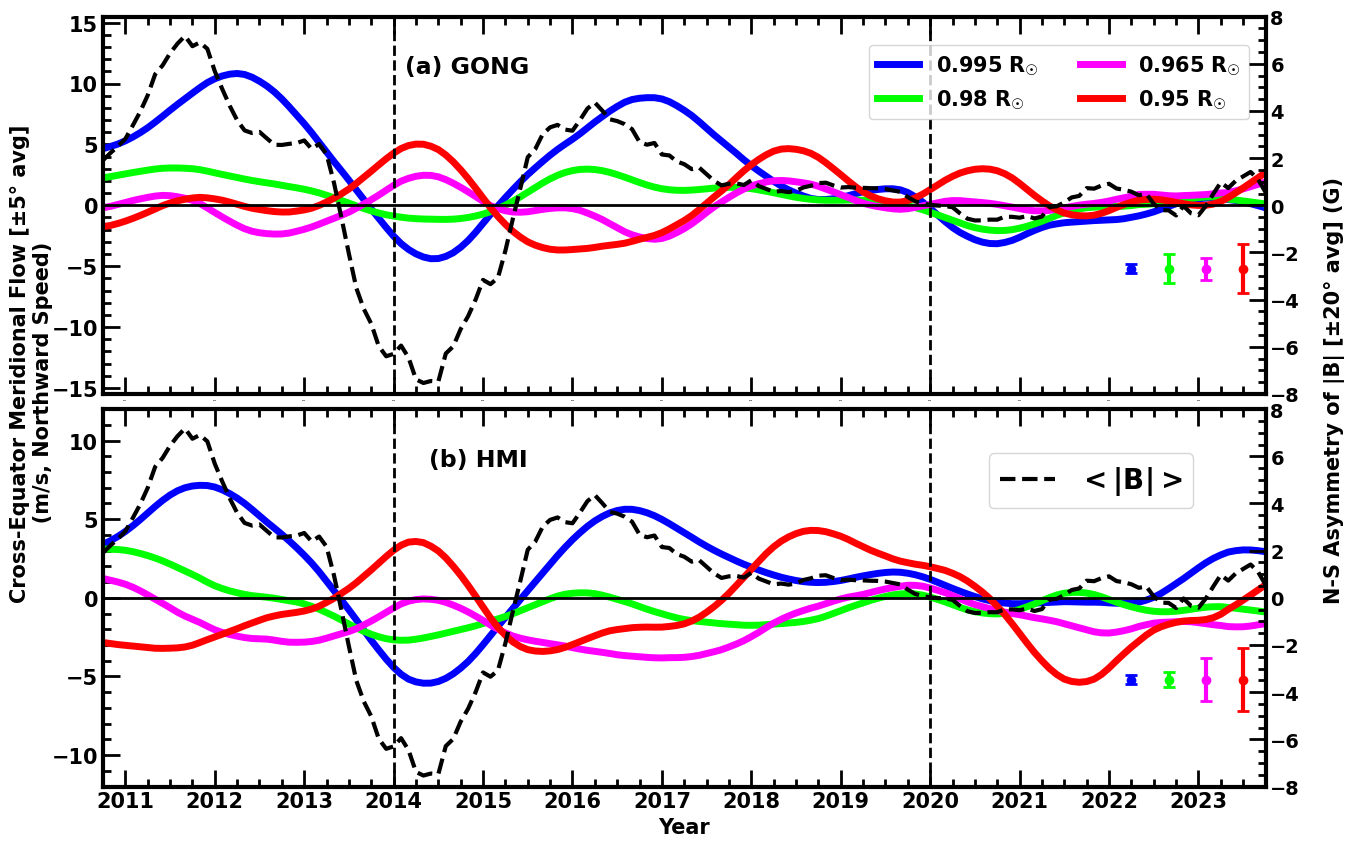}{0.8\textwidth}{}          }

\caption{Cuts across different depths of the 2D time-depth profile of cross-equatorial flows shown in Figure \ref{fig:1} for 
GONG and HMI: blue (0.995 $R_{\sun}$), green (0.98 $R_{\sun}$), magenta (0.965 $R_{\sun}$), and red (0.95 $R_{\sun}$). 
The estimated error for each depth is marked within the panels. The dashed black curve in both panels is the same as in Figure 
\ref{fig:1}, except that here we have applied a 12-month running average, and it corresponds to the right Y-axis.}

\label{fig:2}
\end{figure*}

Positive hemispheric asymmetry (B$_{asym}$ $>$ 0), in our convention, corresponds to the northern hemisphere being more active 
than the southern hemisphere and vice versa. The horizontal dashed line is the zero level for hemispheric asymmetry (right 
Y-axis). Firstly, we confirm the earlier result \citep{komm2022subsurface} that the cross-equatorial flow changes in response 
to the hemispheric asymmetry in magnetic flux: a higher level of northern hemispheric flux during the rising phase of Cycle 24 
(2011 to 2013) caused north-directed near-surface flow across the equator, an opposite situation during the maximum phase 
(2013 to 2015) and again back to north-directed near-surface flow in the declining phase (2016 to 2018). The agreement between 
GONG \& HMI measurements are striking, both in terms of magnitudes and lifetimes of flow structures: a period of about 2 years 
and a magnitude of about 8 m$s^{-1}$. Most interestingly, we find that these flow structures correspond to cross-equatorial 
circulation cells with return flows roughly at depths below 0.97 $R_{\sun}$. To better illustrate the flow amplitudes and 
their changes over depth, we plot in Figure \ref{fig:2} cuts across different depths of the 2D time-depth profile of 
cross-equatorial flows shown in Figure \ref{fig:1}: blue (0.995 $R_{\sun}$), green (0.98 $R_{\sun}$), magenta (0.965 
$R_{\sun}$), and red (0.95 $R_{\sun}$). The error bars shown in Figure \ref{fig:2} are derived from an 
analysis of inverted flow velocities obtained by repeating the inversion 1000 times, with travel times randomly perturbed 
using the estimated uncertainties in the observed values \citep{rajaguru2015meridional}. This estimated error for each depth 
is marked within the panels. The dashed black curve in both panels is the same as in Figure \ref{fig:1}, and corresponds to the 
right Y-axis. In contrast to Figure \ref{fig:1}, a 
12-month running average has been applied to the dashed black curve (right y-axis) to compare with the flow cut (left y-axis), 
which is averaged over the same period. It is clear that the flow profile at the upper layer is strongly positively correlated 
with the hemispheric magnetic asymmetry, whereas at the deeper layer within the NSSL, it is anti-correlated, with the 
changeover happening near 0.965-0.97 $R_{\sun}$. Between 0.97 $R_{\sun}$ and 0.94 $R_{\sun}$, the flow profile keeps its 
direction, remaining negatively correlated with the hemispheric absolute magnetic asymmetry.

\begin{figure*}[!htbp]
\gridline{\fig{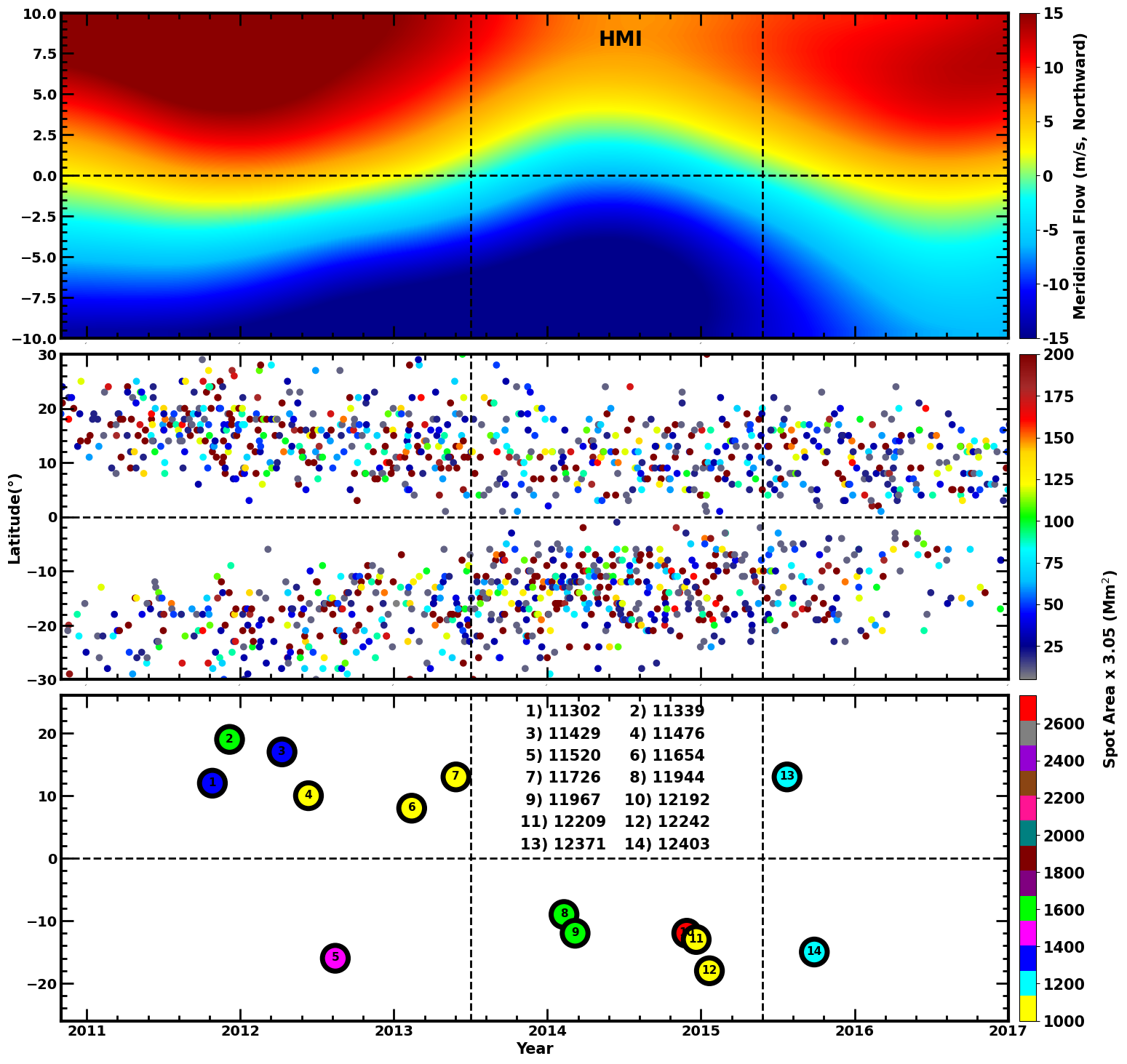}{0.8\textwidth}{}          }

\caption{Time - latitude profile of near-surface (0.99 $R_{\odot}$) meridional flow (over $\pm$10$^{\circ}$ across the 
equator) is compared with that of sunspot locations and sizes. The time period chosen (2011 - 2017) is that of the active 
phase of Solar Cycle 24. The middle panel displays the locations and sizes of all sunspots, with the colorbar restricted to 
200 millionths of a hemisphere ($\mu$HS), while the lower panel shows that of large sunspots with an area greater than 1000 
millionths of a solar hemisphere. The NOAA sunspot group numbers of the large spots are listed within the panel. The two 
vertical dashed lines mark the time of change in flow direction across the equator.}

\label{fig:3}
\end{figure*}

\subsection{Connections Between Cross-Equatorial Flows and Flux Plumes}
\label{subsec: res2.1}

During active phases, especially during the maximum phase of a solar cycle, there are cross-equatorial exchanges of magnetic 
flux, which have been analysed by \citet{2013A&A...557A.141C} and \citet{bisoi2020new}. But neither study had knowledge of 
cross-equatorial meridional flows as a fairly long-lived phenomenon, whose existence was reported in later studies. Moreover, 
our findings here that these flows are part of circulation cells warrant a closer look at the connections between flux plumes 
and flows. However, \citet{2013A&A...557A.141C} did address the combined roles of flux emergence, advection and diffusion in 
the transport of magnetic flux across the equator, and, in particular, demonstrated that cross-equatorial flux plumes 
constitute sudden injections of flux that cannot be explained as a diffusion process. As shown in Figure~\ref{fig:3}, the 
largest sunspot groups (bottom panel) that drive cross-equatorial flows typically emerge away from the equator, while only a 
few smaller groups appear close to the equator (middle panel), some of which deviate from Joy’s law. During cycle 24, however, 
we find that most flux plumes followed the leading polarity of the source hemisphere. This indicates that, in this case, the 
cross-equatorial flux plumes are not attributable to emergence or diffusion processes; instead, advection appears to be the 
dominant mechanism responsible for their occurrence. To address this further, in this Section, we probe the connections 
between the cross-equatorial flows and flux plumes.

The time variation of cross-equatorial flux plumes, B$_{eq}$, is captured by the overplotted pink curve in Figure \ref{fig:1}. 
For solar cycle 24, the leading polarity of the southern hemisphere was positive (red) (see bottom panel of Figure 
\ref{fig:1}). During the maximum of this cycle (2013 - 2015), the southern hemisphere was more active (B$_{asym} <$ 0) while 
the cross-equatorial flux plumes are dominantly positive, {\em i.e.,} the leading polarity flux of the southern hemisphere is 
transported northward across the equator. During 2011–2013, when the northern hemisphere was more active (B$_{asym} >$ 0), the 
flux plumes were negative, corresponding to leading polarity flux from the northern hemisphere crossing the equator. A similar 
episode of negative flux plumes across the equator happened during 2016 - 2017 (when B$_{asym} >$ 0). These observations, 
thus, show that the cross-equatorial surface flows and magnetic flux plumes are in opposite directions. However, our findings 
that cross-equatorial flows are part of circulation cells, which have returning flows at depths below 0.97 $R_{\sun}$, 
indicate that the flux plumes actually are dragged by these outflows at the deeper layers of the NSSL, where the active region 
magnetic flux is rooted. In summary, hemispheric magnetic asymmetry enhances inflows toward the more active hemisphere in the 
upper layer, while outflows in the deeper layer drag the magnetic flux plumes along with them, in the direction opposite to 
the surface inflows.

\begin{figure*}[!htbp]
\gridline{\fig{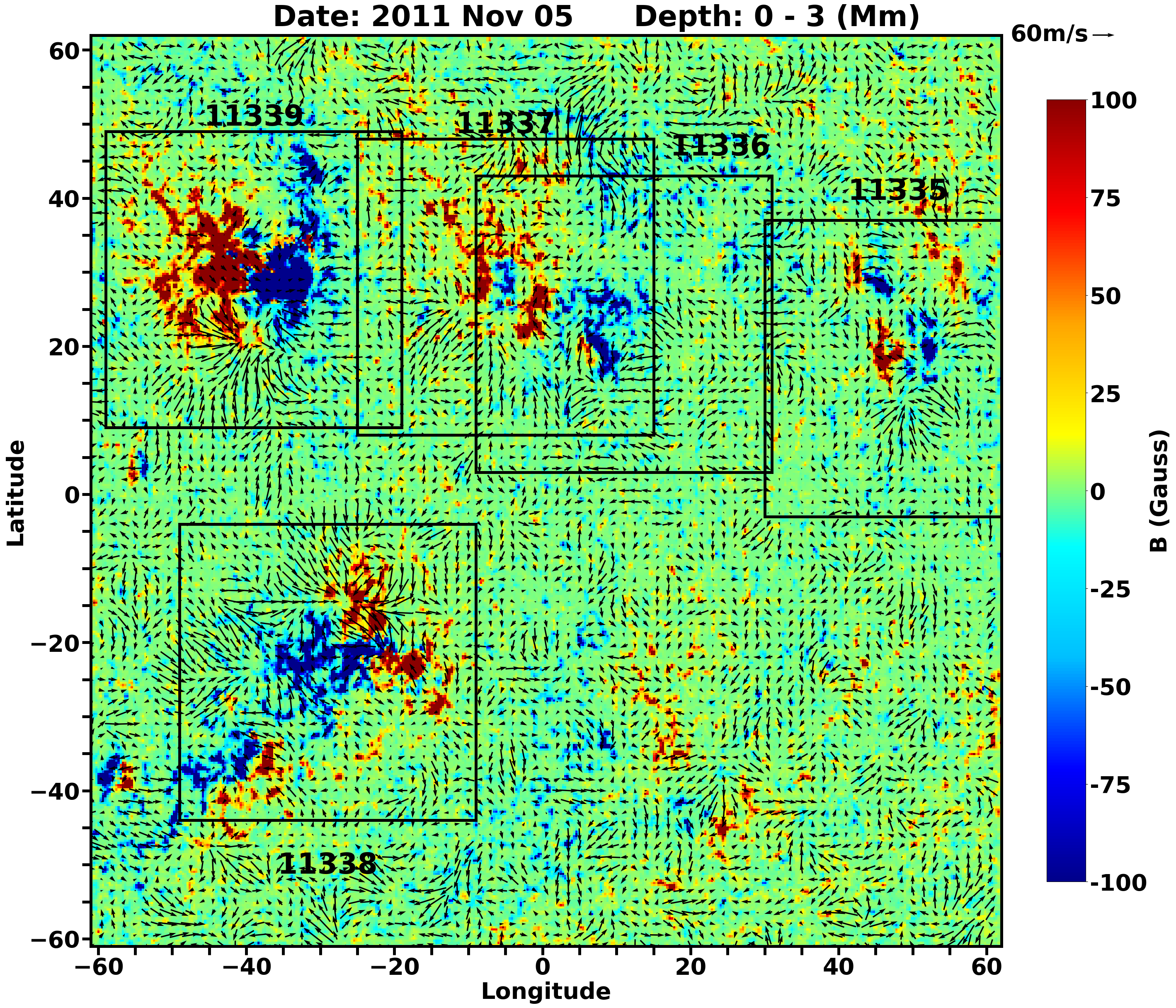}{0.65\textwidth}{}}
\vspace{-20pt} 
\gridline{\fig{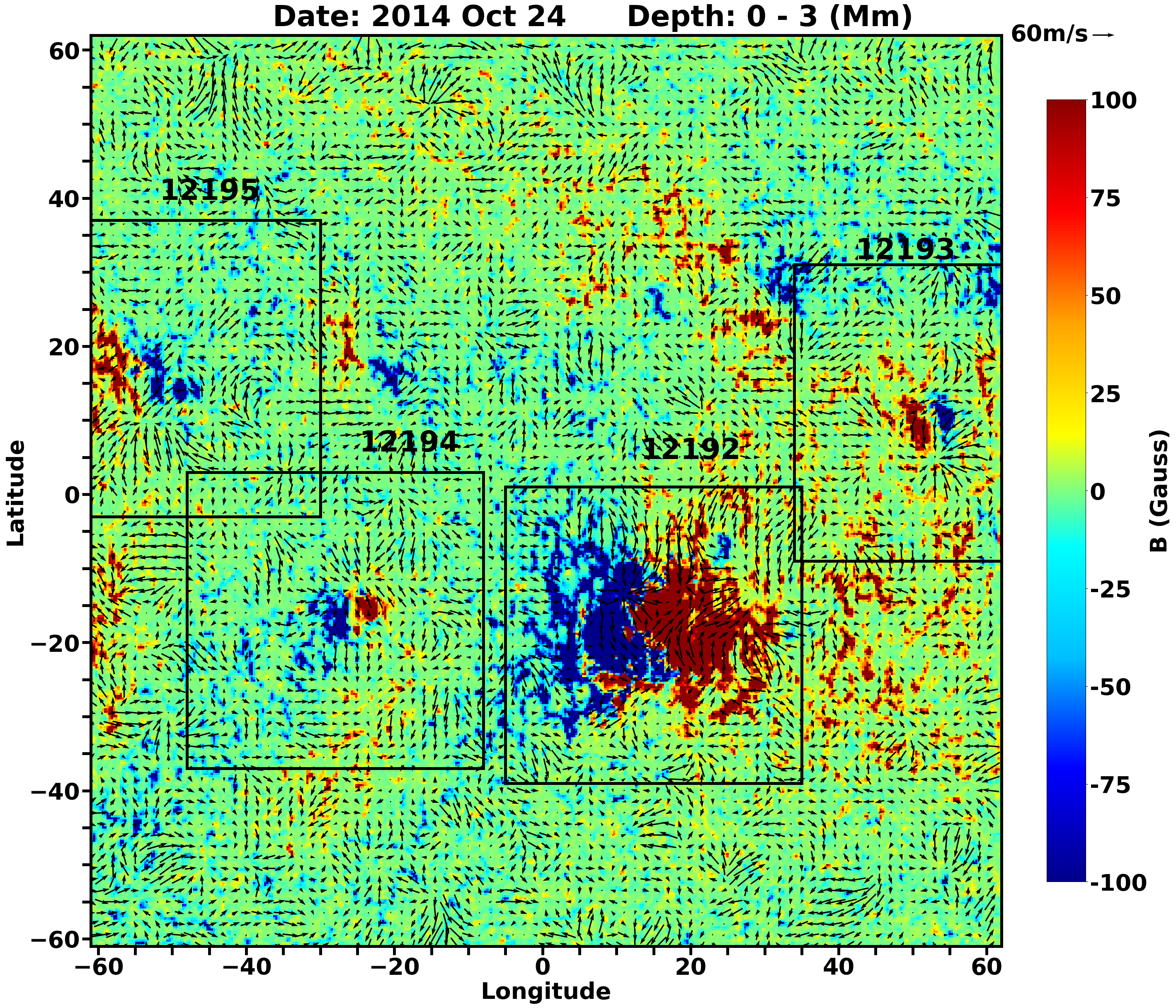}{0.65\textwidth}{}}

\caption{Examples depicting how sunspot regions and their surroundings are removed in flow maps before longitudinally 
averaging them to test the extent of contributions from flows around active regions to the cross-equatorial flows. This test 
is performed on the local 3D (latitude, longitude and depth) time-distance helioseismic flow inversions that map the upper 20 
Mm of the convection zone. The two panels correspond to dates when episodes of large cross-equatorial flows were measured.}

\label{fig:4}
\end{figure*}

\subsection{Sunspots and Cross-Equatorial Flows}
\label{subsec: res2.2}

In this section, we attempt a detailed analysis of how the locations and areas of sunspots are linked to cross-equatorial 
flows. We explore the extent to which the distribution and size of sunspots influence the strength and direction of these 
flows, shedding light on the underlying mechanisms that govern their interaction. The top panel of Figure \ref{fig:3} shows 
the zoomed-in view of the cross-equatorial flow during the cycle maxima (Oct 2010 - Dec 2016) at a depth of 0.995$R{\sun}$. 
Two vertical dashed black lines indicate the flow direction change across the equator. From October 2010 to May 2013, the flow 
was directed toward the northern hemisphere. Between June 2013 and May 2015, it shifted toward the southern hemisphere. From 
June 2015 to October 2023, the flow remained directed northward. In the middle panel of Figure \ref{fig:3}, we present the 
locations of sunspots over the same period mentioned above. The spot area is indicated on the color bar, with a restricted 
upper limit of 200 millionths of the hemisphere, equivalent to 200 × 3.05 ($Mm^{2}$).

\begin{figure*}[!htbp]
\gridline{\fig{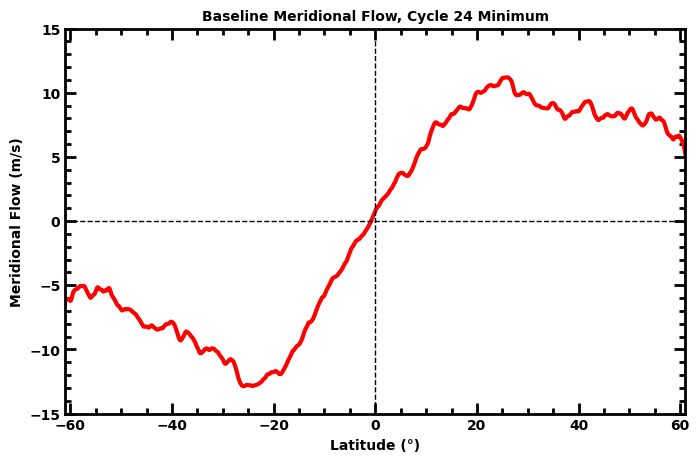}{0.8\textwidth}{} }

\caption{Baseline quiet-Sun meridional flow profile at a depth of 3.5 Mm determined by averaging over a 1-year period during Cycle 24 minimum (mid-2019 to mid-2020).}

\label{fig:5}
\end{figure*}

When the flow is directed toward the northern hemisphere, sunspots in the southern hemisphere become more widely dispersed 
over time, as the flow carries away a significant portion of the magnetic flux. In contrast, sunspots in the northern 
hemisphere tend to remain more clustered. A similar pattern is observed during southward-directed flow, with the roles 
reversed. \citep{komm2022subsurface} reported that during Solar Cycle 23, the cross-equatorial was predominantly directed 
toward the southern hemisphere, whereas in Solar Cycle 24, it was mostly directed northward. Our analysis aligns with previous 
findings, confirming that during Solar Cycle 24, the southward flow persisted for approximately two years (from June 2013 to 
May 2015). However, for the majority of the observed period, the flow was directed northward. We also identified a 
contributing factor: throughout the cycle, the distance between the activity belt in the northern hemisphere and the equator 
remained significantly smaller than in the southern hemisphere. Between October 2010 and December 2016, a total of 1,492 NOAA 
sunspot groups were recorded. From these, we selected the top 1\% of the largest groups, specifically those with an area 
exceeding 1,000 millionths of the hemisphere. Based on this criterion, we identified 14 sunspot groups.

In the lower panel of Figure \ref{fig:3}, we have plotted the largest sunspot groups, with their areas indicated in the color 
bar and group numbers labelled within the figure. Notably, sunspots with areas exceeding 1,000 millionths of the hemisphere 
appear in the photosphere approximately 8 to 12 months after the onset of cross-equatorial flow in the upper layers. For 
instance, during the northward flow that began in October 2010, the first large sunspot group emerged in September 2011 (AR: 
11302, Date: 2011-09-25, Latitude: 12°N, Area: 1300 $\mu$HS). When the cross-equatorial flow was directed southward, the first 
major sunspot group appeared in January 2014 (AR: 11944, Date: 2014-01-09, Latitude: 9°S, Area: 1560 $\mu$HS), following the 
onset of the southward flow in June 2013. We also found that sunspots with areas exceeding 1,000 millionths of the hemisphere 
predominantly form within a latitude range of 8–25 degrees in both hemispheres. The presence of larger sunspots near the 
equator was likely the primary driver of the cross-equatorial flow. A greater number of large sunspots were observed in the 
northern hemisphere when the flow was directed northward, while more large sunspots appeared in the southern hemisphere when 
the flow was directed southward.

\begin{figure*}[!htbp]
\gridline{\fig{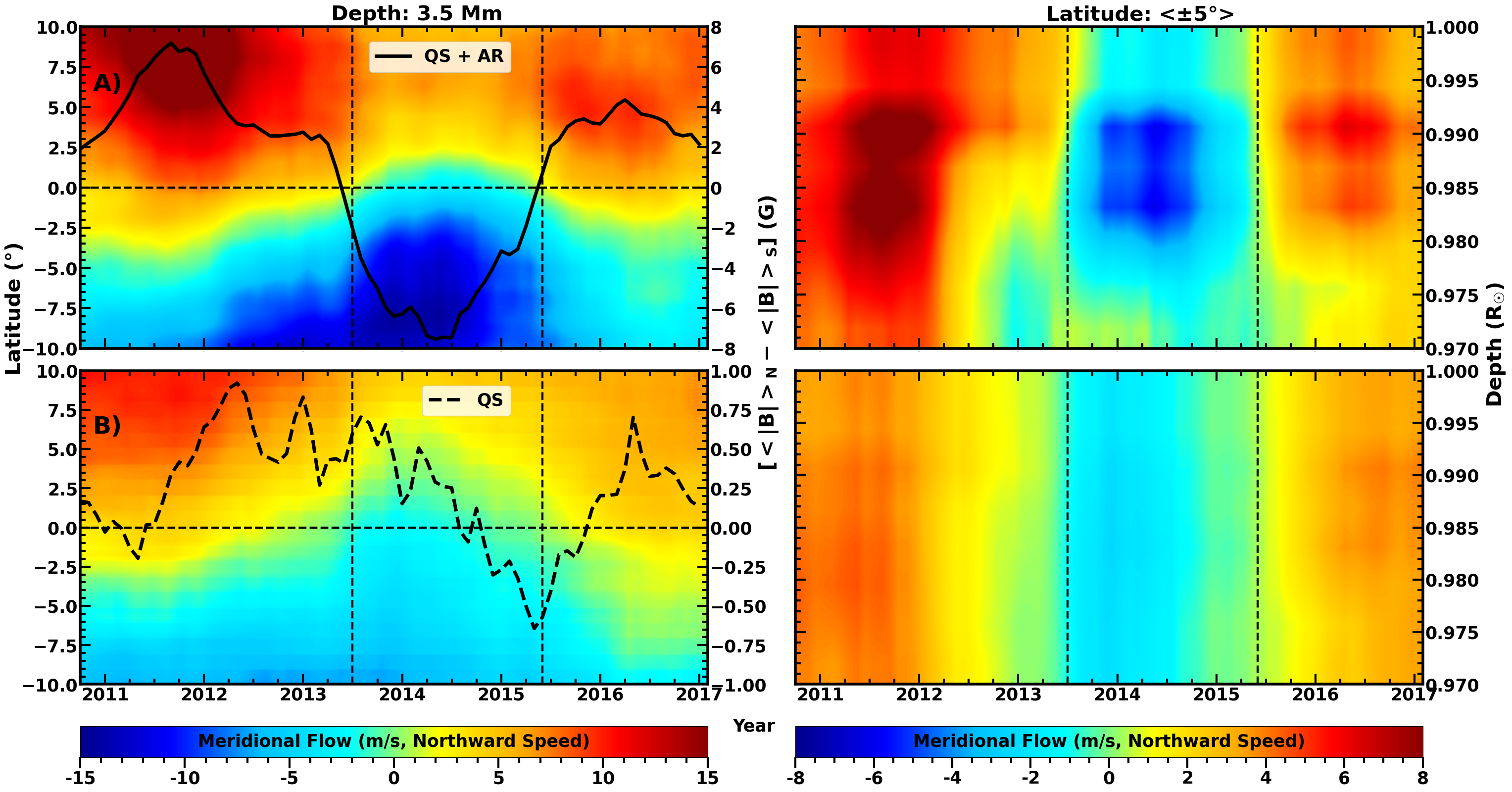}{1\textwidth}{} }

\caption{Comparison of cross-equatorial flow structures before (upper panels) and after removing active regions and their 
surroundings (40$^{\circ}\times$40$^{\circ}$ area) (lower panels). These results are from the local 3D (latitude, longitude 
and depth) time-distance helioseismic flow inversions that map the upper 20 Mm of the convection zone, and the chosen time 
period (2011 - 2017) covers the active phase of Solar Cycle 24. Time - latitude profiles of the meridional flow at 
a near-surface depth (of 3.5 Mm, corresponding to the average over the first three depth points at 1, 0.995, and 
0.99 $R_{\sun}$) are in the left column, and the time-depth profiles covering 0.97 to 1 $R_{\odot}$ are 
in the right one. The two black curves (solid and dashed) represent the asymmetry in absolute magnetic fields with and without active regions, and the two vertical dashed black lines are the same as in Fig:\ref{fig:3}. }

\label{fig:6}
\end{figure*}

\begin{figure*}[!htbp] 
\gridline{\fig{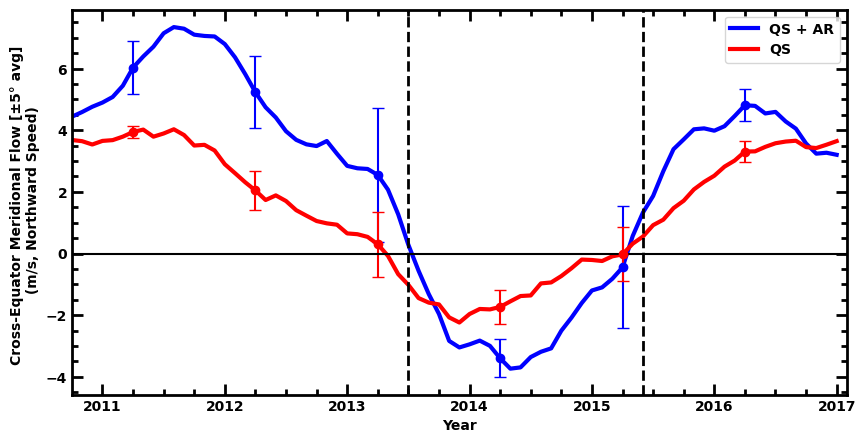}{0.8\textwidth}{} } 

\caption{Comparison of cross-equatorial meridional flow profiles (at a depth of 3.5 Mm) before (blue, $\it{i.e.}$ the whole Sun) and after removing 
(red) the active regions (see text for details). Error bars here correspond to the standard errors determined from the standard
deviation of individual flow measurements over 1-year bins.}

\label{fig:7}
\end{figure*}

%
%

\subsection{Tests Removing Active Region Contributions to Cross-Equatorial Flows}
\label{subsec: res2.3}

Questions on the level of contributions from flows around active regions to variations in large-scale meridional flows are 
still not answered satisfactorily, as there are contradicting results from different analyses 
\citep{hindman2009subsurface,braun2019flows, poulier2022contribution,mahajan2023removal}. Here, we attempt an investigation 
into the extent of active region contributions to cross-equatorial flows by employing the 3D (latitude, longitude, depth) 
local time-distance helioseismic inversions for horizontal velocity fields available in the JSOC time-distance (TD) 
helioseismology pipeline \citep{2012SoPh..275..375Z}, which facilitates deriving longitudinally averaged cross-equatorial 
flows after removing active regions and their surroundings. The chosen time period is October 2010 to December 2016. Firstly, 
we remove the center-to-limb systematics \citep{zhao2012systematic,2018ApJ...853..161C} in these full-disk flow maps, stacked 
from 30$\times$30$^{o}$ tiles, and the large-scale time-averaged background rotation signal following 
\citet{mahajan2023removal}. 

We then derive longitudinally averaged meridional flows for each Carrington rotation before and 
after removing the active regions and their surroundings, as illustrated in Figure \ref{fig:4}. 
We excluded a total of 1,492 NOAA sunspot groups observed between October 2010 and December 2016, covering their entire
lifetimes. For each group, all occurrences were removed within 40°$\times$40° windows, 
centered on the corresponding active region flux centroid retrieved from the Space-Weather HMI Active Region Patches 
(SHARPs; \citet{2014SoPh..289.3549B}) database. During the cycle maximum period, however, such active region removal 
leads to situations of running out of data points or with very few points being averaged (see also \citet{mahajan2023removal}) 
resulting in artificially enhanced flow values relative to the case of full average. To remedy 
this situation, we estimate a baseline quiet-Sun meridional flow profile, over latitude and depth, by averaging over 1 1-year 
period during Cycle 24 minimum (mid-2019 to mid-2020; there still were a few small active regions, and we masked them out) 
and assign it to all identified active region locations that are set for removal. This procedure ensures consistency in the pixel 
statistics before and after the removal of active regions, as the total number of pixels averaged remains the same. The estimated
baseline profile is shown in Figure \ref{fig:5}, which compares well with that determined by \citet{mahajan2023removal}.

\begin{figure*}[!htbp] 
\gridline{\fig{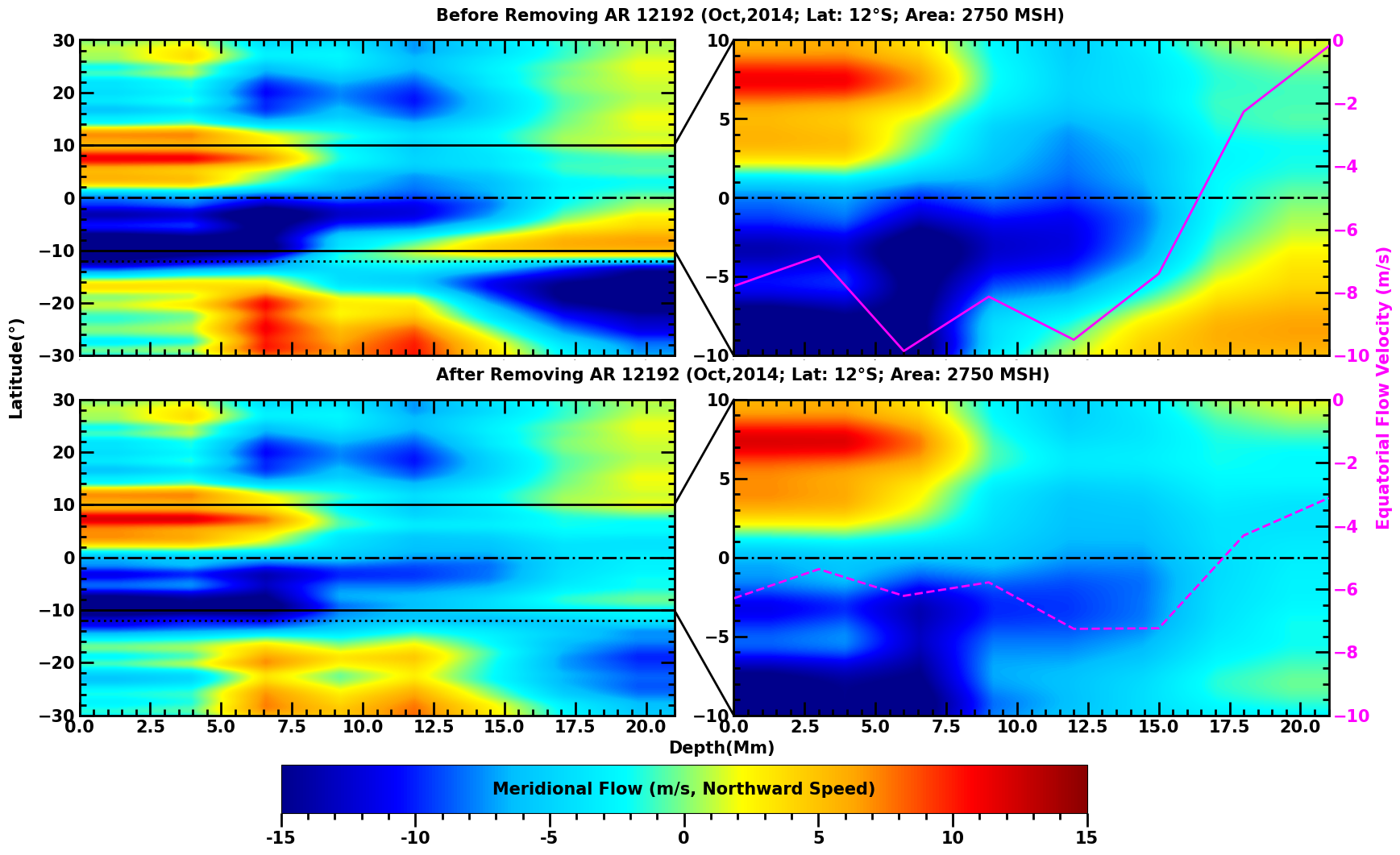}{1\textwidth}{} } 

\caption{One-month averaged flow maps over 0-21 Mm depth, shown before (top panels) and after (bottom panels), removing the 
largest active region (AR 12192). In the left panels, the dotted line marks the latitude of the active region and the dashed 
line marks the equator. The right panels show the corresponding zoomed-in region (10°S-10°N), with Equatorial flow velocities 
overplotted in magenta on the right y-axis.}

\label{fig:8}
\end{figure*}

We show the results of the above exercise in Figure \ref{fig:6}: the temporal variation of the cross-equatorial 
flow at the near-surface depth of 3.5 Mm (average over the first three depths at 1, 0.995 and 0.99 $R_{\sun}$) in the top left 
panel and the depth profile of the same near the equator in the top right panel. Here, a 12-month running average 
has been applied to compare with the Figure \ref{fig:1}; the lower panels show flows recovered after the active regions and 
their surroundings have been removed and replaced by baseline (quiet-Sun) flows as explained above.
Comparing the upper and lower panels, we find that, although the cross-equatorial flow signal has decreased significantly 
after removing the active regions, a residual flow pattern remains. Similarly, the two black curves (solid and dashed) in 
Figure \ref{fig:6}, which represent the asymmetry in absolute magnetic fields with and without active regions, reveal that the 
asymmetry pattern also persists after removing the active regions, albeit with a slight temporal offset. This perhaps 
suggests that when sunspots decay into network and internetwork fields \citep{sen2023dynamics}, the residual magnetic flux 
continues to maintain a north-south asymmetry, contributing to the cross-equatorial flow. Removal of active regions
reduced the asymmetry by about $\pm 7 G$, with a residual asymmetry roughly at $\pm 1 G$ level. 
To make a further quantitative comparison of flow profiles before and after the 
removal of active regions, we plot in Figure \ref{fig:7} the near-surface (average of first three depths covering 0 - 7 Mm)
cross-equatorial flow averaged between 5°S and 5°N depths. The profiles prior to and after active-region removal are shown in 
blue and red, respectively, with error bars representing the standard deviation of measurements within a 1-year bin. Depths up 
to 7 Mm were considered because the strongest inflows occur within the  3.5 - 7 Mm range (see 
Figure \ref{fig:8}). The comparison indicates that although the flow amplitude decreases markedly after the removal of active 
regions, it does not vanish entirely. We discuss this further and other possible causes in Section \ref{sec:conclusion}.

To further test the contributions of active regions, we examine flow structures averaged over individual CRs (or months). 
Among the 14 largest sunspot groups plotted in the bottom panel of Figure \ref{fig:3}, we selected the NOAA sunspot group AR 
12192 (Date: 2014-10-27, Latitude: 12°S, Area: 2750 $\mu$HS), as it was the largest spot during cycle 24. Using local 
time-distance flow maps again, we examine the flow structure across the different depths throughout the full month of its 
occurrence. Figure \ref{fig:8} presents a detailed analysis of the cross-equatorial flow during October 2014. The top panels 
of Figure \ref{fig:8} show the one-month averaged latitude-depth flow profile before removing the active region, while the 
bottom panels present the profile after removing the largest spot, AR 12192. The x-axis represents depth (0 - 21 Mm), while 
the y-axis indicates latitude. The left panels cover the range from 30°S to 30°N, and the right panels provide a zoomed-in 
view spanning 10°S to 10°N. In the left panels, the latitude of the largest sunspot group is marked with a horizontal dotted 
black line. The right y-axis of the right panels shows the meridional flow velocity at the equator (latitude 0°), depicted by 
the magenta curve. During October 2014, the cross equatorial flow was negative, indicating a flow directed towards the 
southern hemisphere. Figure \ref{fig:8} reveals that the cross-equatorial flow primarily originates from deeper layers in the 
opposite hemisphere, extending over a broad latitudinal range compared to the surface flow. In October 2014, when the flow was 
directed southward, a southward (-ve) flow was observed in the deeper layers of the northern hemisphere, extending up to 
latitudes of 0° - 10°N. We also observed inflows toward active region 12192, as indicated by the black dotted horizontal lines 
at 12°S. Notably, inflows originating from higher latitudes and moving toward active regions exhibit maximum amplitude within 
the 5 - 7.5 Mm depth range, while their amplitude significantly decreases in the shallower 0 - 5 Mm layer. This suggests that 
a substantial portion of these inflows originates from deeper layers within the 5 - 7.5 Mm range. Additionally, equatorial 
inflows, those approaching active regions from the equatorial side, display greater amplitude than inflows from higher 
latitudes. In contrast, outflows become prominent at depths of approximately 12 Mm and extend into deeper layers, with 
stronger amplitudes at higher latitudes. Meanwhile, outflows directed toward the equator exhibit lower amplitude compared to 
those occurring at higher latitudes. The strength of the equatorial outflow also depends on the area of the sunspot group.

The bottom panels of Figure \ref{fig:8} show the same flow profiles as the top panels, but after removing the largest sunspot 
group, AR 12192. A direct comparison reveals substantial changes in the flow patterns linked to this active region. In 
particular, the magnitude of inflows toward the active region decreases notably after its removal. The strong equatorial 
outflow associated with AR 12192, clearly visible in the top panels, disappears entirely in the bottom panels, highlighting a 
major alteration in the large-scale flow structure. To quantify this change, we examine the cross-equatorial flow shown by the 
magenta curves in the right panels, with the solid line representing the flow before removal and the dashed line after. These 
curves, plotted on the right y-axis, demonstrate a reduction of about 3 m/s following the exclusion of AR 12192.

\section{Discussion and Conclusion} 
\label{sec:conclusion}

Using time-distance helioseismic measurements of global-scale meridional flows \citep{2025ApJ...984L...1S}, 
we have examined the cross-equatorial part of these flows to depths down to 0.94$R_{\sun}$, covering the whole of the NSSL
over a period of 14 years starting from May 2010. Measuring such flows should be free of leakage of solar rotation caused 
by error in the position angle (P-angle) due to instrumental misalignment. For example, a southward flow signal of about 5 m/s 
across the equator was detected in early time-distance helioseismic studies using data from the Michelson Doppler Imager (MDI) onboard
SOHO was identified as due to a small misalignment of about 0.2$^{\circ}$ in the orientation of the MDI instrument 
\citep{giles1997subsurface,giles2000time,2010Sci...327.1350H,2017A&A...601A..46L}. As indicated in Section \ref{sec:intro},
both the HMI and GONG datasets used here have undergone rigorous alignment verifications employing planetary transits, ensuring
reliable measurement of meridional flow variations near the equator. Combining MDI, GONG and HMI data \citet{komm2022subsurface} 
performed ring diagram analyses of meridional flows covering Cycles 23 and 24 and the early part of Cycle 25, and established 
that meridional flows have cross-equatorial excursions, which, at depths shallower than 10 Mm, are directed toward the hemisphere 
with stronger magnetic flux. The same author also identified them as driven by inflows toward active regions.
Here, we have confirmed and extended earlier findings. During Cycle 24, enhanced magnetic 
activity in the southern hemisphere systematically drove near-surface flows across the equator in the southward direction, 
with amplitudes of about 8 m/s and a lifetime of about 2 years. A striking agreement is observed between GONG and HMI results. 
Importantly, our measurements have imaged the whole of NSSL, revealing a flow reversal around 0.97 $R_{\sun}$.
Flows in the upper NSSL are positively correlated with magnetic asymmetry, while anti-correlated with it beneath 0.97 $R_{\sun}$, 
forming cross-equatorial circulation cells with return flows at deeper layers. These results highlight a 
dynamic coupling between surface magnetic asymmetry and meridional flow structures throughout the NSSL, with indications
that the deeper outflows away from active latitudes play dominant roles in the driving of circulation as well as in the transport of magnetic flux at the surface layers (see below).

Magnetic flux transport generally occurs through a combination of emergence, advection, and diffusion. 
\citet{2013A&A...557A.141C} had, in fact, examined these processes and concluded that cross-equatorial flux plumes largely 
represent sudden, non-diffusive injections of flux. In our observations here, during the maximum phase of solar cycle 24, most 
flux plumes were carried from the more active southern hemisphere toward the northern one. Notably, these plumes were 
transported in the direction opposite to the near-surface cross-equatorial flows. This apparent contradiction can be 
reconciled by considering circulation cells: near-surface inflows converge toward the more active hemisphere, while at deeper 
layers (below 0.97 $R_{\sun}$), return flows or outflows drag the flux plumes across the equator in the opposite direction. 
also note that the deep-rooted magnetic flux is more passively advected at depths where high plasma $\beta$ conditions 
prevail, and hence the deep-layer advection emerges as the dominant mechanism governing the transport of cross-equatorial flux 
plumes that are observed at the surface.

The contribution of active regions to large-scale meridional flow variations remains debated, with earlier studies offering 
conflicting results. Using 3D time-distance helioseismic inversions from the JSOC pipeline (2010–2016), we investigated 
cross-equatorial flows before and after removing active regions and their surroundings. Although the amplitude of the flow 
decreases after removal, the overall flow pattern and magnetic asymmetry remain. 
Removal of active regions reduces the asymmetry by $\pm 7 G$, but a residual asymmetry of $\pm 1 G$ persists, suggesting that 
decayed sunspots to network and internetwork fields maintain the asymmetry. 
During cycle maximum, it becomes particularly challenging to eliminate all magnetic regions, especially those 
associated with pre-emergence and post-emergence phases. \citet{2021A&A...652A.148G} reported that near-surface converging flows of 
approximately 20-30 m/s become detectable about one day prior to flux emergence. Similarly, we may expect active region flows 
to persist for a while after the sunspot flux bundle submerges or disperses. 
In our analysis, the removal of active regions has not covered the above pre-emergence and post-decay or submergence periods
of active regions. The omission of these phases may explain the persisting cross-equatorial flows after the removal of active regions. 
A more detailed investigation of the pre-emergence and decay phases is essential to better constrain and reduce this effect.

We may also speculate on the possibility of magnetic fields confined within the near-surface shear layer (NSSL) but have
not fully emerged to the photosphere. Such hidden flux, though not directly visible at the surface, can still play a 
significant role and contribute to the observed effects.
Our tests removing active region areas further imply that the primary driver of cross-equatorial flows may lie in deeper layers; outflows associated with magnetic flux accumulation at the base of the NSSL could be the 
real source. Importantly, such deeper outflows may exist even 
in the absence of corresponding emerged flux-forming sunspots above them, meaning their contribution cannot be eliminated 
simply by masking active regions at the surface. These deeper outflows can still drive circulation cells whose surface 
manifestations produce cross-equatorial flows, thereby maintaining the observed pattern.

\section{Acknowledgements} 

The HMI data used are courtesy of NASA/SDO and the HMI science teams. Data preparation and processing have utilised the Data 
Record Management System (DRMS) software at the Joint Science Operations Center (JSOC) for NASA/SDO at Stanford University. 
Our sincere thanks go to H.M. Antia for supporting A.S. in performing the inversions. The GONG data used are obtained by the 
NSO Integrated Synoptic Program, managed by the National Solar Observatory, which is operated by the Association of 
Universities for Research in Astronomy (AURA), Inc., under a cooperative agreement with the National Science Foundation and 
with contributions from the National Oceanic and Atmospheric Administration. This work has received funding from the NASA 
DRIVE Science Center COFFIES Phase II CAN 80NSSC22M0162 to Stanford University. S.K. was partly supported by NASA grants 
80NSSC21K0735, 80NSSC19K0261 and 80NSSC20K0194 to NSO. Data-intensive computations in this work have utilised the 
High-Performance Computing facility at the Indian Institute of Astrophysics. A.S. is supported by INSPIRE Fellowship from the 
Department of Science and Technology (DST), Government of India. S.P.R. acknowledges support from the Science and Engineering 
Research Board (SERB, Government of India) grant CRG/2019/003786. We thank the anonymous referee for a thorough review and valuable suggestions that led to significant improvements and additions to the manuscript.

\bibliography{msbib}{}
\bibliographystyle{aasjournal}
\end{document}